\documentclass[letterpaper,reprint,10pt,aps,physrev,superscriptaddress,floatfix]{revtex4-2}
\usepackage[utf8]{inputenc}
\usepackage{bm}
\usepackage[usenames]{color}
\usepackage[dvipsnames]{xcolor}
\usepackage{multirow}
\usepackage{amssymb}
\usepackage{amsbsy}
\usepackage{amsmath}
\usepackage{stmaryrd}
\usepackage{graphicx}
\usepackage{epsfig}
\usepackage{placeins}
\usepackage[normalem]{ulem}
\usepackage{bbold}
\usepackage{braket}
\usepackage{blindtext}

\usepackage{xcolor}
\definecolor{midnight3}{HTML}{4a6d88}

\usepackage[colorlinks,linkcolor=midnight3,citecolor=midnight3,urlcolor=midnight3]{hyperref}
\usepackage{filecontents}

\newcommand{\BOne}{\text{B}_1}
\newcommand{\BTwo}{\text{B}_2}
\newcommand{\B}{\text{B}_1}

\usepackage{scalerel}
\usepackage{tikz}
\usetikzlibrary{calc}
\usetikzlibrary{patterns}
\usetikzlibrary{svg.path}
\definecolor{orcidlogocol}{HTML}{A6CE39}
\tikzset{
  orcidlogo/.pic={
    \fill[orcidlogocol] svg{M256,128c0,70.7-57.3,128-128,128C57.3,256,0,198.7,0,128C0,57.3,57.3,0,128,0C198.7,0,256,57.3,256,128z};
    \fill[white] svg{M86.3,186.2H70.9V79.1h15.4v48.4V186.2z}
                 svg{M108.9,79.1h41.6c39.6,0,57,28.3,57,53.6c0,27.5-21.5,53.6-56.8,53.6h-41.8V79.1z M124.3,172.4h24.5c34.9,0,42.9-26.5,42.9-39.7c0-21.5-13.7-39.7-43.7-39.7h-23.7V172.4z}
                 svg{M88.7,56.8c0,5.5-4.5,10.1-10.1,10.1c-5.6,0-10.1-4.6-10.1-10.1c0-5.6,4.5-10.1,10.1-10.1C84.2,46.7,88.7,51.3,88.7,56.8z};
  }
}

\newcommand\orcid[1]{\!%
  \href{https://orcid.org/#1}{%
    \mbox{%
      \scaleto{%
        \begin{tikzpicture}[yscale=-1,transform shape]
          \pic{orcidlogo};
        \end{tikzpicture}
      }{8pt}%
    }%
  }%
}

\makeatletter
\newcommand*{\balancecolsandclearpage}{
  \close@column@grid
  \clearpage
  \twocolumngrid
}
\makeatother

\makeatletter
\def\maketitle{
\@author@finish
\title@column\titleblock@produce
\suppressfloats[t]
\let\and\relax
\let\affiliation\@gobble@opt@one
\let\address\affiliation
\let\author\@gobble
\@author@init
\let\@authors\@empty
\let\@authors@curr\@empty
\let\@affil@list\@empty
\let\keywords\@gobble
\let\@keywords\@empty
\let\email\@gobble
\let\@address\@empty
\let\thanks\@gobble
\titlepage@sw{ %
\clearpage
}{}%
}
\makeatother

\begin{document}

\title{Spin-1/2 XXZ chain coupled to two Lindblad baths: \\
Constructing nonequilibrium steady states from equilibrium correlation 
functions}

\author{Tjark Heitmann~\orcid{0000-0001-7728-0133}}
\email{tjark.heitmann@uos.de}
\affiliation{Department of Mathematics/Computer Science/Physics,
University of Osnabr\"uck, D-49076 Osnabr\"uck, Germany}

\author{Jonas Richter~\orcid{0000-0003-2184-5275}}
\affiliation{Department of Physics, Stanford University, 
Stanford, California 94305, USA}
\affiliation{Institut f\"ur Theoretische Physik, Leibniz 
Universit\"at Hannover, 30167 Hannover, Germany}

\author{Fengping Jin~\orcid{0000-0003-3476-524X}}
\affiliation{Institute for Advanced Simulation, Jülich Supercomputing Centre, 
Forschungszentrum Jülich, D-52425 Jülich, Germany}

\author{Sourav Nandy}
\affiliation{Jo\v{z}ef Stefan Institute, SI-1000 Ljubljana, Slovenia}

\author{Zala Lenar\v{c}i\v{c}~\orcid{0000-0001-8374-8011}}
\affiliation{Jo\v{z}ef Stefan Institute, SI-1000 Ljubljana, Slovenia}

\author{Jacek Herbrych~\orcid{0000-0001-9860-2146}}
\affiliation{Wroclaw University of Science and Technology, 50-370 Wroclaw,  
Poland}

\author{Kristel Michielsen~\orcid{0000-0003-1444-4262}}
\affiliation{Institute for Advanced Simulation, Jülich Supercomputing Centre, 
Forschungszentrum Jülich, D-52425 Jülich, Germany}

\author{Hans De Raedt~\orcid{0000-0001-8461-4015}}
\affiliation{Zernike Institute for Advanced Materials, University of Groningen, 
NL-9747 AG Groningen, Netherlands}

\author{Jochen Gemmer}
\affiliation{Department of Mathematics/Computer Science/Physics,
University of Osnabr\"uck, D-49076 Osnabr\"uck, Germany}

\author{Robin Steinigeweg~\orcid{0000-0003-0608-0884}}
\email{rsteinig@uos.de}
\affiliation{Department of Mathematics/Computer Science/Physics,
University of Osnabr\"uck, D-49076 Osnabr\"uck, Germany}

\date{\today}

\begin{abstract}
State-of-the-art approaches to extract transport coefficients of 
many-body quantum systems broadly fall into two categories: (i) they target the 
linear-response 
regime in terms of equilibrium 
correlation functions of the closed system; or (ii) 
they consider an open-system situation typically modeled by a Lindblad 
equation, where a nonequilibrium steady state emerges from driving the 
system at its boundaries. While quantitative agreement between (i) and 
(ii) has been found for selected model and parameter choices, also disagreement 
has been pointed out in the literature.
Studying magnetization transport in the spin-$1/2$ XXZ chain, we here 
demonstrate that at weak driving, the nonequilibrium steady state in an open
system, including its buildup in time, can remarkably be constructed just 
on the basis of correlation functions in the closed system. We numerically 
illustrate this direct correspondence of closed-system and open-system 
dynamics, and show that it allows the treatment of comparatively large 
open systems, 
usually only accessible to matrix product state simulations. We also point out 
potential pitfalls when extracting transport coefficients from nonequilibrium 
steady states in finite systems. 
\end{abstract}
\maketitle

{\it Introduction.} 
Our understanding of the properties of many-body quantum 
systems out of equilibrium has seen remarkable advances in the last decades 
thanks to various experimental and theoretical breakthroughs 
\cite{Bloch2008, Abanin2019, Polkovnikov2011, Eisert2015, Dalessio2016}.
Central questions are concerned with the emergence of particular (thermal or 
nonthermal) steady states in the long-time limit, but
also with the (universal) properties of the actual nonequilibrium process 
towards such states in the course of time \cite{Abanin2019, Polkovnikov2011, 
Eisert2015, 
Dalessio2016}. 
Broadly speaking, these and related questions are usually studied in two 
different scenarios: (i) the system of interest is perfectly isolated from its 
environment and evolves unitarily in time; (ii) the system's time evolution is 
nonunitary due to an explicit coupling to an external bath which can affect 
the dynamics (see, e.g., Ref.\ \cite{Breuer2007, Lange2018, Rubio-Abadal2019}). 

In systems with a global conservation law, a fundamental role is 
played by transport processes \cite{Bertini2021}. Quantum 
transport is also a prime example of a research question that is explored 
both 
from 
a closed-system and an open-system perspective.  
In closed systems, a widely used approach is linear response theory, where the 
Kubo formula allows for the extraction of 
transport coefficients from equilibrium 
correlation functions, which can be studied 
in the time or
frequency domain and in real or momentum space \cite{Bertini2021}. 
While nonintegrable systems are
expected to exhibit normal diffusion 
\cite{Lux2014, Bohrdt2017, Richter2018a}, 
the concrete calculation of diffusion
constants for specific models turns 
out to be a hard task in practice.
This difficulty has been one of the motivations for the development of
sophisticated numerical methods 
\cite{Long2003, Heidrich-Meisner2007, Grossjohann2010, Karrasch2012, 
Elsayed2013, Steinigeweg2014, Leviatan2017,Rakovszky2018, White2018, Wurtz2018, 
Richter2019b, Heitmann2020, Ye2020, Jin2021, Rakovszky2022}.
Moreover, some classes of models can generically
feature anomalous subdiffusion or superdiffusion in certain 
parameter regimes 
\cite{Nandkishore2015, Ljubotina2017, Luitz2017a, Abanin2019, 
Gopalakrishnan2019a, Kloss2019,
Ljubotina2019, Heitmann2020a, Schuckert2020, 
Bulchandani2021, Serbyn2021, Singh2021, Nandy2022,
Richter2022a, Richter2022}.

In contrast, when studying transport in an open-system setting, the model of 
interest is often 
coupled at its edges to two reservoirs, e.g., at different temperatures or 
chemical potentials, leading to a nonequilibrium
steady state in the long-time limit. Then, the profile and currsnt of this
steady state yield information on the transport behavior
\cite{Michel2003, Wichterich2007, Znidaric2011, Znidaric2016}. 
A popular description of such an open system is provided by 
the Lindblad quantum master equation \cite{Breuer2007}, not least since it 
allows for efficient numerical simulations based on matrix product states, 
giving access to comparatively large system sizes 
\cite{Prosen2009, Verstraete2004, Zwolak2004, Weimer2021, Lenarcic2020, 
Prelovsek2022, Nandy2022}. 
While quantitative agreement of transport coefficients according to the 
Lindblad description with those from closed-system approaches 
has been found for selected models and parameter regimes 
\cite{Steinigeweg2009a, Znidaric2018, Heitmann2022}, 
also disagreement has been pointed out in the
literature \cite{Purkayastha2018}, and there is no proof that both approaches 
have to agree 
\cite{Kundu2009, Purkayastha2018, Purkayastha2019, Znidaric2019, Bertini2021}. 

From a physical perspective, computed transport coefficients for 
a given system 
should of course be 
independent of the method 
employed. 
In fact, some of us have recently shown that the dynamics of closed and 
open systems can be connected with each 
other in a certain simple setting. Specifically, Ref.\ \cite{Heitmann2022} 
considered 
an initially homogeneous system coupled locally to a single Lindblad bath, which 
induces a net magnetization into the system. Remarkably, it was shown 
that if the 
Lindblad driving is weak, the 
flow of the magnetization in the open system, i.e., the broadening of 
the nonequilibrium density profile, can be described by an appropriate 
superposition of equilibrium correlation functions in the closed 
system. Building on this result, we here go beyond Ref.\ 
\cite{Heitmann2022} in a crucial point and explore the more common situation of 
two 
Lindblad baths inducing a nonequilibrium steady state. Considering 
magnetization transport in the paradigmatic {spin-$1/2$} XXZ chain as an 
example, we demonstrate that the steady state in the open system 
can be constructed 
on the basis of correlation functions in the closed system. We 
support our analytical results by large-scale numerical simulations and 
show that our scheme enables efficient unravelings of Lindblad 
equations for systems with up to $36$  sites, which are usually only accessible 
with matrix product state techniques. 

{\it Closed System.} 
We consider the one-dimensional XXZ model, which is described 
by the Hamiltonian
\begin{equation}
H = J \sum_{r=1}^N (S_r^x S_{r+1}^x + S_r^y S_{r+1}^y + \Delta S_r^z S_{r+1}^z) 
\, ,
\end{equation}
where $S_r^j$ (${j = x, y, z}$) are spin-1/2 operators at site $r$, ${J > 0}$
is the antiferromagnetic coupling constant, and $\Delta$ denotes the
anisotropy in the $z$ direction. Moreover, $N$ is the number of
sites and we employ periodic boundary conditions, ${S_{N+1}^j \equiv S_1^j}$. 
The XXZ chain conserves the global magnetization, $[H,\sum_r S_r^z] = 0$, and 
we will particularly focus on the regime ${\Delta > 1}$, where it is 
well-established that spin transport is diffusive \cite{Bertini2021}. 
This diffusive transport behavior can, for instance, be seen in the Gaussian 
shape of 
the infinite-temperature spin-spin correlation function at $\Delta = 1.5$
\cite{Steinigeweg2017a}, see 
Fig.\ \ref{Fig::Trans_Closed}(a), 
\begin{equation}\label{Eq::Korrel}
 \langle S_r^z(t) S_{r'}^z(0) \rangle_\text{eq} = \frac{\text{tr}[e^{iHt} S_r^z 
e^{-iHt}S_{r'}^z]}{2^N}\ .
\end{equation}
The root-mean-squared displacement of the above grows as 
${\Sigma(t) \propto \sqrt{t}}$, see 
Fig.~\ref{Fig::Trans_Closed}(b), where 
${\Sigma^2(t) = \sum_r (r-r')^2 C_{rr'}(t) - [\sum_r (r-r') C_{rr'}(t)]^2}$ 
and ${C_{rr'}(t) = 4\langle S_r^z(t) S_{r'}^z\rangle_\text{eq}}$. 
Moreover, a diffusion coefficient can be defined as 
${2D(t) = \tfrac{d}{dt}\Sigma^2(t)}$ \cite{Steinigeweg2009b}. 
As shown in Fig.~\ref{Fig::Trans_Closed}(b), $D(t)$ 
takes on a constant value ${D/J \approx 0.6}$ for ${tJ \lesssim 10}$, 
which is approximately independent of time 
(and system size \cite{Steinigeweg2015,Steinigeweg2017a}) 
and consistent with other results in the literature 
\cite{Steinigeweg2009c,Steinigeweg2011,Karrasch2014,Prelovsek2022}.

In the following, we will show that the equilibrium correlation function
$\langle S_r^z(t) S_{r'}^z(0) \rangle_\text{eq}$ in Eq.~\eqref{Eq::Korrel} is
not only central to transport in the closed system, but can remarkably be used
to predict the buildup of a nonequilibrium steady state in an open-system
situation where the spin chain is weakly driven by two Lindblad baths. While we
focus on the integrable XXZ chain as a concrete example
due to its interesting transport properties, we expect our conceptual findings
to apply to a wider range of models. In particular, while our derivation
\cite{Heitmann2022,SupplementalMaterial} is largely model-independent, it
implicitely assumes sufficiently fast local equilibration, which should be even
better fulfilled in nonintegrable chaotic systems.
\begin{figure}[tb]
\includegraphics[width=0.90\columnwidth]{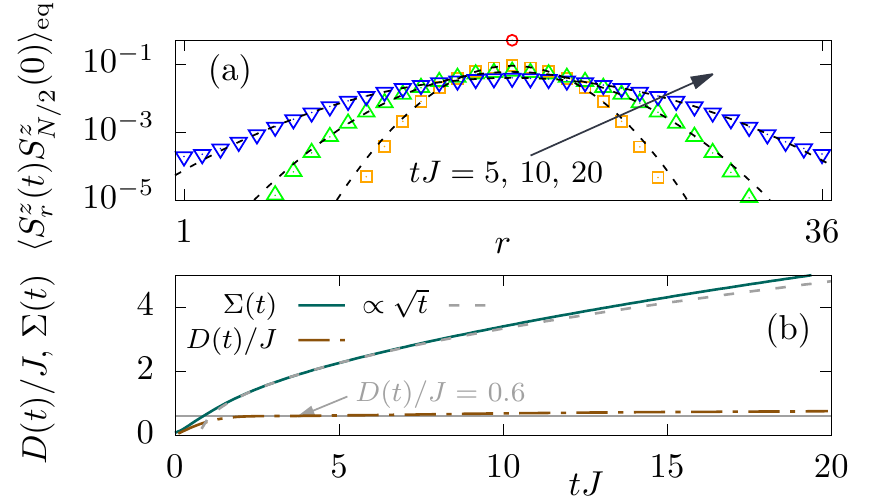}
\caption{(a) Infinite-temperature correlation function $\langle 
S_r^z(t)S_{N/2}^z(0)\rangle_\text{eq}$ for ${N = 36}$ and ${\Delta = 1.5}$. The 
dashed curves indicate Gaussians, see also \cite{Steinigeweg2017a}. 
(b) Diffusive growth of root-mean-squared 
displacement ${\Sigma(t) \propto \sqrt{t}}$. A diffusion constant 
${D/J \approx 0.6}$ can be extracted 
from the approximately constant 
plateau of ${2D(t) = \frac{d}{dt}\Sigma^2(t)}$ at ${tJ \lesssim 10}$. 
For longer times, finite-size effects become relevant.}
\label{Fig::Trans_Closed}
\end{figure}

{\it Open System.} 
Let us consider a scenario, where the XXZ chain is explicitly coupled to an 
environment. We describe this 
setting with a Lindblad equation, 
\begin{equation}\label{Lind1}
\dot{\rho}(t) = {\cal L} \, \rho(t)  = i [\rho(t),H] + {\cal D} \, \rho(t) \, ,
\end{equation}
which consists of a coherent time evolution of the density matrix $\rho$ 
with respect to $H$ and an incoherent 
damping term, 
\begin{equation}
{\cal D} \, \rho(t) = \sum_j \alpha_j \Big ( L_j \rho(t) L^\dagger_j - 
\frac{1}{2} \{ \rho(t), L_j^\dagger L_j \} \Big )\ , 
\end{equation}
with non-negative rates $\alpha_j$, Lindblad operators $L_j$, and the 
anticommutator $\{ \bullet, \bullet \}$. While the derivation of this equation 
can be a subtle task for a given microscopic model 
\cite{Wichterich2007, DeRaedt2017}, it is the most general form of a 
time-local quantum master equation, which maps any density matrix 
to a density matrix, i.e., which preserves trace, 
hermiticity, and positivity
\cite{Breuer2007}. 
Here, we choose \cite{Bertini2021}
\begin{eqnarray}\label{eq:jump_operators}
L_1 = S_{\BOne}^+ \, , \quad && \alpha_1 = \gamma (1 + \mu) \\
L_2 = L_1^\dagger = S_{\BOne}^- \, , \quad && \alpha_2 = \gamma (1 - \mu) 
\,  \\
L_3 = S_{\BTwo}^+ \, , \quad && \alpha_3 = \gamma (1 - \mu) \\
L_4 = L_3^\dagger = S_{\BTwo}^- \, , \quad && \alpha_4 = \gamma (1 + 
\mu) 
\, \label{Jumplast} ,
\end{eqnarray}
where $\gamma$ is the system-bath coupling and $\mu$ is the driving strength. 
$L_1$ and $L_2$ are local Lindblad operators at site $\BOne$ and flip a 
spin up and down, respectively. $L_3$ and $L_4$ act similarly on another site 
$\BTwo$. In the following, we set ${\BOne = 1}$ and ${\BTwo = N/2+1}$. Note 
that we still consider periodic boundary conditions. 
However, our approach can also generally be 
applied to open boundaries with the two baths at the system's edges 
${\BOne = 1}$ and ${\BTwo = N}$, and we present results for this setting in 
\cite{SupplementalMaterial}. 
For ${\mu > 0}$, 
the first (second) bath induces a net polarization of $\mu/2$ ($-\mu/2$), 
leading to a steady state in the long-time limit with a 
characteristic density profile and a constant current. 
Note that, while the Lindblad modeling \eqref{Lind1} -
\eqref{Jumplast} is standard in the context of transport in quantum lattice
models \cite{Bertini2021}, there exist other approaches to open-system dynamics
which can also address potential non-Markovian effects \cite{DeVega2017}.
\begin{figure}[t]
\centering
\includegraphics[width=0.90\columnwidth]{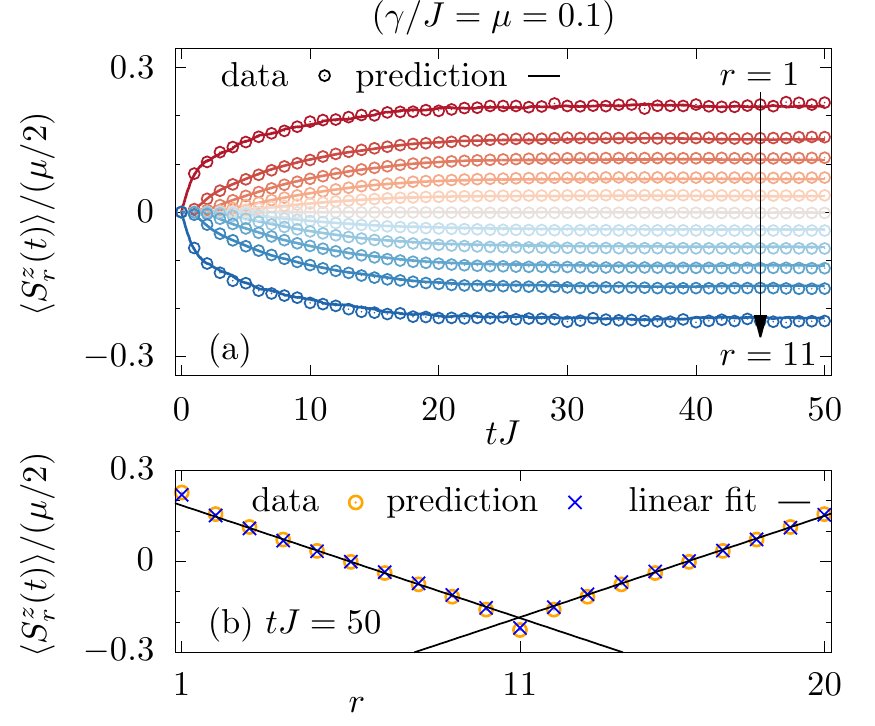}
\caption{Open-system dynamics for the spin-$1/2$ XXZ chain coupled to two 
Lindblad baths, as obtained for anisotropy ${\Delta = 1.5}$, ${N = 20}$ sites
(with periodic boundary conditions), small coupling ${\gamma / J = 0.1}$, and  
weak driving ${\mu = 0.1}$. Numerical 
results from the full stochastic 
unraveling (data) are compared to the prediction 
based on closed-system correlation functions [cf. Eq.\ \eqref{Eq::Average}]. 
(a) Time evolution of the local
magnetization $\langle S_r^z(t) \rangle$ for different sites $r$.
(b) Site dependence of the steady state at ${t J = 50}$. 
}
\label{fig:comparison}
\end{figure}

In addition to the long-time limit, we are 
interested in the temporal buildup of
the steady state. Thus, we study the time evolution of 
local densities
\begin{equation}\label{Dyn::Open}
\langle S_r^z(t) \rangle = \text{tr}[\rho(t) S_r^z] \, ,
\end{equation}
which depends on the parameters of the system $H$, but also on the bath 
parameters $\gamma$ and $\mu$. As an initial state, we here consider a 
homogeneous situation with ${\rho(0) \propto \mathbb{1}}$ being the 
infinite-temperature ensemble. 

{\it Quantum-trajectory approach.} One possibility to solve the Lindblad 
equation is 
given by the concept of stochastic unraveling, which relies on pure states $| 
\psi \rangle$ rather than density matrices \cite{Dalibard1992, Michel2008}. 
It consists of an alternating 
sequence of stochastic jumps with one of the Lindblad operators and 
deterministic evolutions governed by an effective Hamiltonian 
${H_\text{eff} = H  -\tfrac{i}{2} \sum_j \alpha_j \, L_j^\dagger L_j}$.
For our choice of Lindblad operators, 
\begin{equation}\label{Eq::HEffFull}
H_\text{eff} = H  - i \gamma + i \gamma \mu (n_{\BOne} - n_{\BTwo})\ , 
\end{equation}
with ${n_r = S_r^+ S_r^- = S_r^z + \mathbb{1}/2}$. 
For weak driving ${\mu \ll 1}$, 
the time scale on which the last term in Eq.\ \ref{Eq::HEffFull}
affects the dynamics is much longer than the typical time scale between jumps.
Thus, the effective
Hamiltonian can be approximated as
\begin{equation}
H_\text{eff} \approx H  - i \gamma
\end{equation}
and the time evolution of a pure state reads
\begin{equation}\label{Eq::Damping}
| \psi(t) \rangle \approx e^{- \gamma t} \, e^{-i H t} \, | \psi(0) \rangle 
\, ,
\end{equation}
i.e., apart from the scalar damping term, the dynamics is generated by the 
closed system $H$ only. 
The approximation in Eq.\ \eqref{Eq::Damping} 
is one of the main 
ingredients to establish a 
correspondence between the dynamics of the 
isolated and the weakly-driven XXZ chain 
below. For larger 
values of $\mu$, the effective Hamiltonian generating the dynamics of 
$|\psi(t)\rangle$ also involves the two 
operators $n_{\BOne}$ and $n_{\BTwo}$, cf.\ Eq.\ \eqref{Eq::HEffFull}.

Naturally, since $H_\text{eff}$ is a non-Hermitian operator, the norm of 
a 
pure state is not conserved as a function of time. As a consequence, for a 
given $\varepsilon$ drawn at random from a uniform distribution ${] 0,1]}$, 
there is a time, where the condition 
${\left\lVert \vert\psi(t) \rangle \right\rVert^2 > \varepsilon}$ is first 
violated. At this time, a jump with one of the Lindblad operators occurs and 
the 
new and normalized pure state reads
\begin{equation}
| \psi'(t) \rangle = 
\frac{L_j | \psi(t) \rangle}{\left\lVert L_j|\psi(t)\rangle \right\rVert} 
\, ,
\end{equation}
where the specific jump is chosen with probability
\begin{equation}
p_j = \frac{\alpha_j\left\lVert L_j|\psi(t)\rangle\right\rVert^2}
{\sum_j\alpha_j \left\lVert L_j | \psi(t) \rangle \right\rVert^2} \, . 
\label{eq:probability}
\end{equation}
After this jump, the next deterministic evolution takes 
place. This sequence of stochastic jumps and deterministic 
evolutions leads to a particular trajectory $| \psi_\text{T}(t) \rangle$.
The time-dependent density matrix according to the Lindblad equation can 
eventually be 
approximated by the average over different trajectories $\mathrm{T}$.
Thus, expectation values read
\begin{equation}
\langle S_r^z(t) \rangle \approx \frac{1}{\text{T}_\text{max}} 
\sum_{\text{T}=1}^{\text{T}_\text{max}} \frac{\langle \psi_\text{T}(t) | S_r^z 
| \psi_\text{T}(t) \rangle}{\left\lVert |\psi_\text{T}(t) \rangle\right\rVert^2} 
\, ,
\end{equation}
where ${\text{T}_\text{max}}$ is the number of trajectories.

In order to mimic the homogeneous state ${\rho(0) \propto \mathbb{1}}$, we use random 
pure states as initial condition for the stochastic unraveling, 
\begin{equation}\label{eq:initial_state}
| \psi(0) \rangle \propto  
\sum_j c_j \, | \phi_j \rangle \, ,
\end{equation}
where the real and imaginary parts of the coefficients $c_j$
in some given basis $| \phi_j \rangle$ 
are drawn at random according to a Gaussian probability distribution 
with zero mean. Crucially, by exploiting the concept of quantum typicality 
\cite{Gemmer2004, Goldstein2006, Popescu2006, Reimann2007, Bartsch2009, 
Steinigeweg2017a, Balz2018}, expectation values ${\bra{\psi}\bullet \ket{\psi}}$ 
of local 
observables evaluated within such random states can be related to 
infinite-temperature averages $\text{tr}[\bullet]/2^N$. This is used in 
the following to connect the
equilibrium correlation functions 
$\langle S_r^z(t) S_{r'}^z(0)\rangle_\text{eq}$ 
[Eq.~\eqref{Eq::Korrel}] to the dynamics $\langle S_r^z(t) \rangle$ in the 
open system [Eq.~\eqref{Dyn::Open}].
\begin{figure}[t]
\centering
\includegraphics[width=0.90\columnwidth]{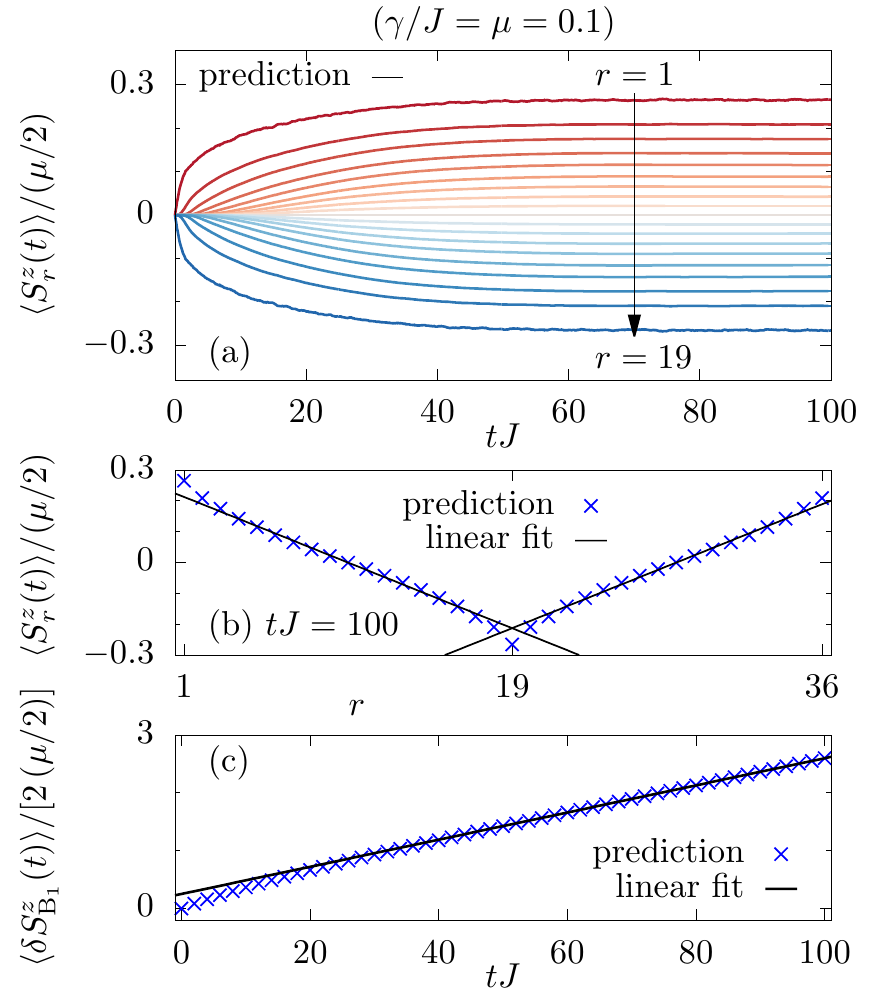}
\caption{[(a) and (b)] Analogous data as in Fig.\ \ref{fig:comparison}, but now 
for ${N=36}$ sites, for which stochastic unraveling is unfeasible.
(c) Magnetization injected by the first bath as a function of 
time, see also \cite{SupplementalMaterial}. A diffusion constant 
${D/J \approx 0.99}$ \cite{D2} can be 
extracted from the slopes in (b) and (c).}
\label{fig:prediction_N36}
\end{figure}

{\it Constructing steady states from correlation functions.} 
In Ref.\ \cite{Heitmann2022}, it was demonstrated that individual 
quantum trajectories of the open system can be described 
by closed-system equilibrium correlation functions if the driving by the 
Lindblad bath is weak. 
We here build on this result and apply it to the case 
of two Lindblad baths leading to a nonequilibrium steady state.
While we relegate details of the derivation to the supplemental 
material \cite{SupplementalMaterial}, we find that for small 
coupling $\gamma$ and weak driving $\mu$, the local magnetization dynamics 
within a single trajectory T can be approximated as 
${d_{r,\text{T}}(t) \approx \bra{\psi_\text{T}(t)}S_r^z 
\ket{\psi_\text{T}(t)}/\left\lVert |\psi_\text{T}(t) \rangle\right\rVert^2}$, 
where
\begin{align}
  d_{r,\text{T}}(t)=2\mu\sum_j &A_j \, \Theta(t -
  \tau_j) \, \mathcal{C}_r(t-\tau_j) 
\label{eq:convolution}
\end{align}
with ${\mathcal{C}_r(t)\equiv\langle S_r^z(t) S_{\BOne}^z\!(0)\rangle_\text{eq} 
- \langle S_r^z(t) S_{\BTwo}^z\!(0)\rangle_\text{eq}}$.
Here, ${\langle \bullet \rangle_\text{eq} = \text{tr}[\bullet]/2^N}$ denotes the 
infinite-temperature ensemble,
$\Theta(t)$ is the Heavyside function, and the sum runs 
over the jump times $\tau_j$ of the particular trajectory $\text{T}$. 
Moreover, the amplitudes 
$A_j$ in Eq.~\eqref{eq:convolution} read 
\begin{align}
A_j &= \frac{a_j - d_{\BOne,\text{T}}(\tau_j - 0^+)}{\mu}\\
\label{eq:amplitudes}
\text{with} \quad
a_j &= \frac{\mu - 2 \, d_{\BOne,\text{T}}(\tau_j - 0^+)}{2 - 4 \mu \, 
d_{\BOne,\text{T}}(\tau_j - 0^+)} \, ,
\end{align}
where ${A_j \to 1/2}$ for ${d_{\BOne,\text{T}}(\tau_j - 0^+) \to 0}$. 
Note that, due to the symmetry 
$d_{\BOne,\text{T}}(\tau_j - 0^+)=-d_{\BTwo,\text{T}}(\tau_j - 0^+)$, 
only $\BOne$ enters the above expressions. 
Equation 
\eqref{eq:convolution} is the main result 
of this Letter. It predicts the magnetization dynamics in the open system by 
suitably superimposing equilibrium correlation functions of the closed 
system involving the two bath sites $\BOne$ and $\BTwo$. In particular, from 
Eq.\ \eqref{eq:convolution}, the trajectory-averaged magnetization 
dynamics follows as 
\begin{equation}\label{Eq::Average}
 \langle S_r^z(t) \rangle \approx \frac{1}{\text{T}_\text{max}} 
\sum_{\text{T}=1}^{\text{T}_\text{max}} d_{r, \text{T}}(t) \, ,
\end{equation}
where each $d_{r, \text{T}}(t)$ is evaluated for a different sequence 
${(\tau_1, \tau_2, \ldots)}$ of $\tau_j$. Given the 
exponential damping in Eq.~\eqref{Eq::Damping}, the $\tau_j$ can be generated 
as ${\tau_{j+1} = \tau_j - \ln \varepsilon_{j+1}/2 \gamma}$, 
where $\varepsilon_{j+1}$ are random numbers drawn from a box 
distribution 
${]0,1]}$. If the correlation functions 
${\langle S_r^z(t) S_{\BOne}^z\!(0) \rangle_\text{eq}}$ and 
${\langle S_r^z(t) S_{\BTwo}^z\!(0) \rangle_\text{eq}}$ are known, 
it is thus straightforward to 
evaluate Eq.\ \eqref{Eq::Average} for a large number of sequences.

{\it Numerical Illustration.} 
We now test our theoretical prediction  
and its accuracy for a specific example, namely 
the spin-$1/2$ XXZ chain 
with ${\Delta = 1.5}$,  
${N = 20}$, and periodic boundary conditions. 
The baths are located at $B_1 = 1$ and $B_2 = 11$ and we 
focus on small 
coupling 
${\gamma/J = 0.1}$ and weak driving ${\mu = 0.1}$.
Additional data for other values of
$\Delta$, $\gamma$, and $\mu$,  
as well as for open boundary conditions 
can be found in \cite{SupplementalMaterial}. 

Our theoretical prediction \eqref{Eq::Average} 
is carried out numerically for ${{\cal O}(10^4-10^5)}$ different
sequences of jump times, which turns out to be 
sufficient to obtain negligibly small statistical errors.
For comparison, we simulate the exact dynamics of the open system 
by performing a stochastic unraveling of the Lindblad equation. 
We stress that 
while 
\eqref{Eq::Average} is derived in the 
limit of weak driving, cf.\ 
Eq.~\eqref{Eq::Damping}, the stochastic unraveling is 
here performed for the 
full $H_\text{eff}$ in Eq.~\eqref{Eq::HEffFull}. 

In 
Fig.~\ref{fig:comparison}, we depict the outcome of the comparison.
In Fig.~\ref{fig:comparison}(a), we show the time evolution of the local
magnetization ${\langle S_r^z(t) \rangle}$ for different sites $r$. The site 
dependence of the steady-state profile is
depicted in Fig.~\ref{fig:comparison}(b) and is 
well described by a linear
function, except for the sites located exactly at the bath contacts.
Importantly, we observe a
remarkably good agreement between our prediction \eqref{Eq::Average} and the 
exact open-system dynamics for all 
times up to ${t J = 50}$, where the steady-state
profile is already established. This confirms our main result 
\eqref{eq:convolution}. 
We also note that for larger values of $\gamma$ and $\mu$, 
deviations are expected to become more 
pronounced, see \cite{SupplementalMaterial}.

For ${N \gg 20}$, 
stochastic unraveling cannot be carried
out, since the required average over many 
trajectories becomes unfeasible. 
In contrast, our theoretical prediction \eqref{Eq::Average} can be evaluated for 
larger system sizes, since only the equilibrium correlation functions are needed
in Eq.\ \eqref{eq:convolution}. In 
particular, by relying on quantum typicality \cite{Jin2021, Heitmann2020}, we 
simulate 
these correlation functions for up to $N = 36$ lattice sites on 
J\"ulich's ``JUWELS'' supercomputer. 
As shown in Figs.~\ref{fig:prediction_N36}(a) and \ref{fig:prediction_N36}(b), 
we are thus able to describe the buildup of a nonequilibrium steady state 
in a $N = 36$ XXZ chain weakly driven by Lindblad baths 
at sites $\BOne = 1$ and $\BTwo = 19$.
Open-system simulations for such system sizes are typically only accessible 
with matrix product state techniques, which are in turn usually restricted to 
open boundary conditions.

{\it On the extraction of transport coefficients.} 
In the nonequilibrium steady state, the diffusion constant can be calculated as
${D = -\langle j_r \rangle/\nabla \langle S_{r} \rangle}$
for some site $r$ in the bulk away from the bath sites.
Here, $j_r$ is the local spin-current operator. Its 
expectation value can be expressed as
${\langle j_r \rangle = \frac{\text{d}}{\text{d}t} \langle \delta 
 S^z_{\BOne}\!(t) \rangle/2}$, where 
 ${\langle \delta S^z_{\BOne}\!(t) \rangle}$ is 
the magnetization injected by the first bath (see 
\cite{SupplementalMaterial} for more details), 
and the 
factor $1/2$ takes into account 
that magnetization can flow to the left and to the 
right of this bath, due to periodic boundary conditions.
As shown in Fig.~\ref{fig:prediction_N36}(c), ${\langle \delta S^z_{\BOne}\!(t) 
\rangle}$ grows linearly in time (i.e., a constant $\langle j_r\rangle$) and we 
can thus 
evaluate the diffusion constant in the steady state as 
the ratio of the slopes in Figs.~\ref{fig:prediction_N36}(b)
and \ref{fig:prediction_N36}(c). 
In this way, we obtain a value ${D/J\approx 0.99}$ 
which differs notably from what we found earlier in 
the context of Fig.~\ref{Fig::Trans_Closed}. 
\begin{figure}[tb]
 \centering
 \includegraphics[width=0.98\columnwidth]{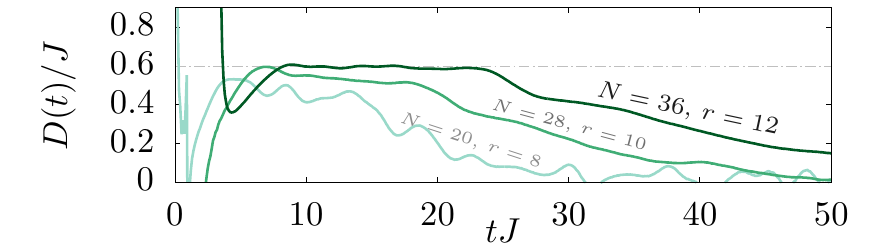}
 \caption{Diffusion coefficient ${D(t) = \partial_t \langle S_r^z(t) 
\rangle/\nabla^2 \langle S_r^z(t) \rangle}$ based on our prediction for 
the dynamics of the weakly-driven open system in 
Fig.\ \ref{fig:prediction_N36}. Data is 
obtained at lattice site $r = 12$ and 
the approximately constant plateau for $tJ\lesssim 25$ 
is consistent with the 
transport behavior of the isolated XXZ chain in Fig.\ \ref{Fig::Trans_Closed}.
Similar data for predictions in smaller systems with $N=20$ and $N=28$ 
are shown for comparison. 
}
 \label{Fig4}
\end{figure}

This discrepancy may be explained by finite-size effects.
Specifically, as also apparent in Fig.~\ref{Fig::Trans_Closed}(a), the
equilibrium correlation function $\langle S_r^z(t) S_{r'}^z(0)
\rangle_\text{eq}$ is affected by finite-size effects already at ${tJ \sim
20}$, where the broadening of the density profile has explored the full system.
These effects then likely translate into the steady state in the
weakly-driven open system and its finite-$N$ estimate of the diffusion
constant. Such finite-size effects demonstrate that care must be taken when
extracting transport properties both in closed and open systems. Importantly,
we stress that the main conceptual result of our work, i.e., establishing a
connection between weakly-driven Lindblad dynamics and closed quantum systems,
remains unabated. In the supplemental material
\cite{SupplementalMaterial}, we provide more details on this issue:
Specifically, one can assume an ideal situation where the closed system behaves
perfectly diffusive without finite-size corrections (in contrast to Fig.\
\ref{Fig::Trans_Closed}), in which case the equilibrium correlation functions
$\langle S_r^z(t) S_{r'}^z(0)\rangle_\text{eq}$ follow analytically as damped
modified Bessel functions \cite{SupplementalMaterial}. Using this idealized
Ansatz, we find that the nonequilibrium steady state indeed yields the same
diffusion constant as the closed system.

We note that one can extract a diffusion coefficient also 
from the finite-time dynamics of the open system, even 
before the steady state is 
established, via 
${D(t)=\partial_t \langle S_r^z(t) \rangle/\nabla^2 \langle S_r^z(t) \rangle}$, 
where ${ \nabla^2 \langle S_r^z(t) \rangle = \langle S_{r-1}^z(t)
\rangle - 2 \langle S_r^z(t) \rangle + \langle S_{r+1}^z(t) 
\rangle}$.   
We are able to find a $D(t)$ in Fig.\ \ref{Fig4} that 
exhibits an approximately constant plateau ${D/J \approx 0.6}$ 
for ${tJ \lesssim 25}$ (while at longer $t$ the behavior becomes 
uncontrolled due to dividing two small numbers), 
consistent with our analysis of the closed system 
in Fig.~\ref{Fig::Trans_Closed}.

{\it Conclusion.} 
Considering the example of magnetization transport in the
spin-$1/2$ XXZ chain, we have connected linear response theory to 
the dynamics in an open quantum system driven by two Lindblad baths. 
Specifically, building on Ref.\ \cite{Heitmann2022}, we have shown 
that, at weak driving, the nonequilibrium steady state and its 
buildup in time can be constructed by 
suitably superimposing equilibrium 
correlation functions of the closed system.

Conceptually, our results for a specific model might reflect 
the natural expectation that transport 
coefficients obtained from closed-system and open-system 
approaches should agree with each other, at least if 
the driving is sufficiently weak. 
While we have presented data for 
systems with periodic boundary 
conditions, we provide 
additional results in \cite{SupplementalMaterial}, where we consider 
the more common case of open boundaries with Lindblad driving at the 
edge spins. In particular, we find that our main result \eqref{Eq::Average} 
works convincingly also in this case and is in good 
agreement with state-of-the-art simulations based on time-evolving block 
decimation \cite{Verstraete2004,Zwolak2004}. 
From a practical perspective, our results enable the treatment of quite large 
open systems, which are usually not accessible by full stochastic unraveling.
It would be an interesting attempt
to generalize our setting to other jump operators, e.g., dephasing 
noise with $L_j = S_j^z$, and other questions beyond quantum transport.

{\it Acknowledgments.} 
We sincerely thank J.~Wang for fruitful discussions. 
Our research has been funded by the Deutsche
Forschungsgemeinschaft (DFG), projects 397107022 (GE 1657/3-2), 397300368 (MI
1772/4-2), and 397067869 (STE 2243/3-2), within DFG Research Unit FOR 2692,
grant no.\ 355031190.
J.\,R.\ acknowledges funding from the European Union's Horizon Europe research 
and innovation programme, Marie Sk\l odowska-Curie grant no.\ 101060162, and the 
Packard Foundation through a Packard Fellowship in Science and Engineering.
We gratefully acknowledge the 
\href{www.gauss-centre.eu}{\color{black}Gauss Centre for Supercomputing e.V.} 
for funding this project by providing computing time on the 
GCS Supercomputer JUWELS \cite{JUWELS} at Jülich Supercomputing Centre (JSC).
Z.\,L.\ and S.\,N.\ acknowledge support by the projects J1-2463 and P1-0044 
program of the Slovenian Research Agency, EU via QuantERA grant T-NiSQ, and also 
computing time for the TEBD calculations at the supercomputer Vega at the 
Institute of Information Science (IZUM) in Maribor, Slovenia.
We also acknowledge computing time at the HPC3 at University Osnabr\"uck, which
has been funded by the DFG, grant no.\ 456666331.

\bibliographystyle{apsrev4-1_titles}
\bibliography{paper.bib,SM.bib}

\balancecolsandclearpage


\title{Supplemental material of\\
``Spin-1/2 XXZ chain coupled to two Lindblad baths: \\
Constructing nonequilibrium steady states from equilibrium correlation 
functions''}
\maketitle

\setcounter{figure}{0}
\setcounter{equation}{0}
\setcounter{page}{1}
\renewcommand{\thefigure}{S\arabic{figure}}
\renewcommand{\theequation}{S\arabic{equation}}

\section*{Details on the derivation of the theoretical prediction}

In this section, we are going to sketch the derivation of our theoretical 
prediction given in Eq.\ \eqref{eq:convolution} of the main text, which is an 
extension of the derivation in Ref.~\cite{Heitmann2022}, where only a single 
Lindblad bath was considered, instead of the two baths treated here.

To this end, let us for the moment consider a simple scenario
featuring a jump with the Lindblad operator $L_1$ immediately at some time 
$t$, say ${t = 0}$. Then,
\begin{equation}
  |\psi(0)\rangle \to |\psi'\rangle \propto L_1\ket{\psi(0)}\, .
\end{equation}
For a random initial state $|\psi(0)\rangle$, as given 
in Eq.~\eqref{eq:initial_state}, this jump results in a random superposition
over a subset of pure states with a spin-up at site $\B$, which mimics 
${\rho(0) \propto \mathbb{1} + S^z_{\B}}$.

At weak driving $\mu \ll 1$, the subsequent deterministic evolution before 
the next jump reads
\begin{equation}\label{Eq::DEVo}
d_r(t) \! \equiv \! \frac{\langle \psi'(t) | S_r^z | \psi'(t) 
\rangle}{\left\lVert| 
\psi'(t) \rangle \right\rVert^2} \! 
\approx \! \langle \psi' | e^{i H t} S_r^z e^{-i H t} | \psi' \rangle \, ,
\end{equation}
cf.\ Eq.~(\ref{Eq::Damping}) of the main text.
Now, using the concept of typicality, Eq.~\eqref{Eq::DEVo}
can be rewritten as 
\begin{equation}\label{Eq::TypEqdr}
\frac{d_r(t)}{2} \approx \langle S_r^z(t)S_{\B}^z\!(0)\rangle_\text{eq}
\end{equation}
with ${S_r^z(t) = e^{i H t} S_r^z e^{-i H t}}$ and
${\langle \bullet \rangle_\text{eq} = \text{tr}[\bullet]/2^N}$ denoting the 
infinite-temperature ensemble \cite{Steinigeweg2017a}. 
Analogously, one can obtain such a relation for the other 
possible jumps with the Lindblad operators $L_j$, which then involve either 
${\langle S_r^z(t)S_{\B}^z\!(0)\rangle_\text{eq}}$ or 
${\langle S_r^z(t)S_{\BTwo}^z\!(0)\rangle_\text{eq}}$. 
Note that in the derivation of Eq.\ \eqref{Eq::TypEqdr}, we
used the facts that $S_r^z = n_r - \mathbb{1}/2$, $(n_r)^2 = n_r$, and 
$\text{tr}[S_r^z(t)] = 0$, see e.g., Ref.\ \cite{Heitmann2020} for more 
details.

For the above homogeneous initial state $| \psi(0) \rangle$, the jump 
probabilities according to Eq.~\eqref{eq:probability} are simply given by $p_j 
= \alpha_j/4\gamma$. Consequently, averaging over all 4 jumps 
possibilities, 
\begin{align}
\frac{\bar{d}_r(t)}{2} =& \, (p_1-p_2)\langle S_r^z(t) S_{\BOne}^z\!(0) 
\rangle_\text{eq} \nonumber \\
&+ (p_3-p_4)\langle S_r^z(t) S_{\BTwo}^z\!(0)
  \rangle_\text{eq}
\end{align}
yields the theoretical prediction
\begin{equation}
 \bar{d}_r(t) =  \mu \langle S_r^z(t) S_{\BOne}^z\!(0)  \rangle_\text{eq} - \mu 
\langle S_r^z(t) S_{\BTwo}^z\!(0)
\rangle_\text{eq}
\end{equation}
for the time evolution after the first and before the second jump.

Let us consider a second jump at a later time $\tau$. The corresponding 
jump probabilities $p_j$ can then be derived based on typicality arguments. 
To this end, we assume a random pure state 
${| \psi(\tau - 0^+) \rangle}$ 
right before the jump with ${\bar{d}_{\BOne}\!(\tau - 0^+) \neq 0}$ 
and ${\bar{d}_{\BTwo}\!(\tau - 0^+) \neq 0}$.
Then, due to the symmetry
\begin{equation}
\bar{d}_{\BOne}\!(\tau - 0^+)=-\bar{d}_{\BTwo}\!(\tau - 0^+)
\end{equation}
we have
\begin{align}
y_1& =\left\lVert L_1|\psi(\tau-0^+)\rangle\right\rVert^2 = 
\frac{1}{2}-\bar{d}_{\BOne}\!(\tau - 0^+) \\
y_2 &=\left\lVert L_2|\psi(\tau-0^+)\rangle\right\rVert^2=
\frac{1}{2}+\bar{d}_{\BOne}\!(\tau - 0^+) \\
y_3&=\left\lVert L_3|\psi(\tau-0^+)\rangle\right\rVert^2=
\frac{1}{2}-\bar{d}_{\BTwo}\!(\tau - 0^+)=y_2 \\
y_4&=\left\lVert L_4|\psi(\tau-0^+)\rangle\right\rVert^2=
\frac{1}{2}+\bar{d}_{\BTwo}\!(\tau - 0^+)=y_1 
\end{align}
with ${y_1 + y_2 + y_3 + y_4 = 1}$. Thus, the jump probabilities 
read
\begin{align}
p_1 = p_4 = \frac{1}{2}\frac{(1+\mu) y_1}{(1+\mu) y_1 + (1-\mu) y_2}\\
p_2 = p_3 = \frac{1}{2}\frac{(1-\mu) y_1}{(1+\mu) y_1 + (1-\mu) y_2}
\end{align}
with $p_1 + p_2 + p_3 + p_4 = 1$. A straightforward calculation yields
\begin{equation}
p_1-p_2 = p_4-p_3 = \frac{\mu - 2 \,\bar{d}_{\B}\!(\tau - 0^+)}{2 - 4 \mu 
\,\bar{d}_{\B}\!(\tau - 0^+)} \, .
\end{equation}
Using this expression, we can define the amplitude
\begin{equation}
A_\tau= \frac{(p_1-p_2)-\bar{d}_{\BOne}(\tau - 0^+)}{\mu}\, ,
\end{equation}
to incorporate both the probabilities for the next jump as well as the fact 
that some magnetization is already induced at the bath site.
With this, we can eventually formulate a theoretical prediction for the 
deterministic 
evolution after the second jump, in analogy to the case of a single Lindblad 
bath \cite{Heitmann2022}. This prediction reads
\begin{align}\label{eq:twojumps}
\bar{d}_r(t) = \mu\, \mathcal{C}_r(t) + 2\mu \, A_{\tau} \, 
\Theta(t-\tau)\, \mathcal{C}_r(t-\tau) \, 
\end{align}
with the Heavyside function $\Theta(t)$ and the abbreviation
\begin{equation}
  \mathcal{C}_r(t) =\langle S_r^z(t) S_{\BOne}^z\!(0)\rangle_\text{eq} 
- \langle S_r^z(t) S_{\BTwo}^z\!(0)\rangle_\text{eq} \, .
\end{equation}
Reiterating this procedure finally yields a 
generalization of Eq.~\eqref{eq:twojumps} to a sequence of jump times $\tau_j$,
\begin{align}
  \bar{d}_{r}(t) = 2\mu\sum_j &A_j \, \Theta(t -
 \tau_j) \, \mathcal{C}_r(t-\tau_j)\, ,
 \end{align}
i.e., Eq.~\eqref{eq:convolution} in the main text.

\begin{figure}[t]
  \centering
  \includegraphics[width=0.90\columnwidth]{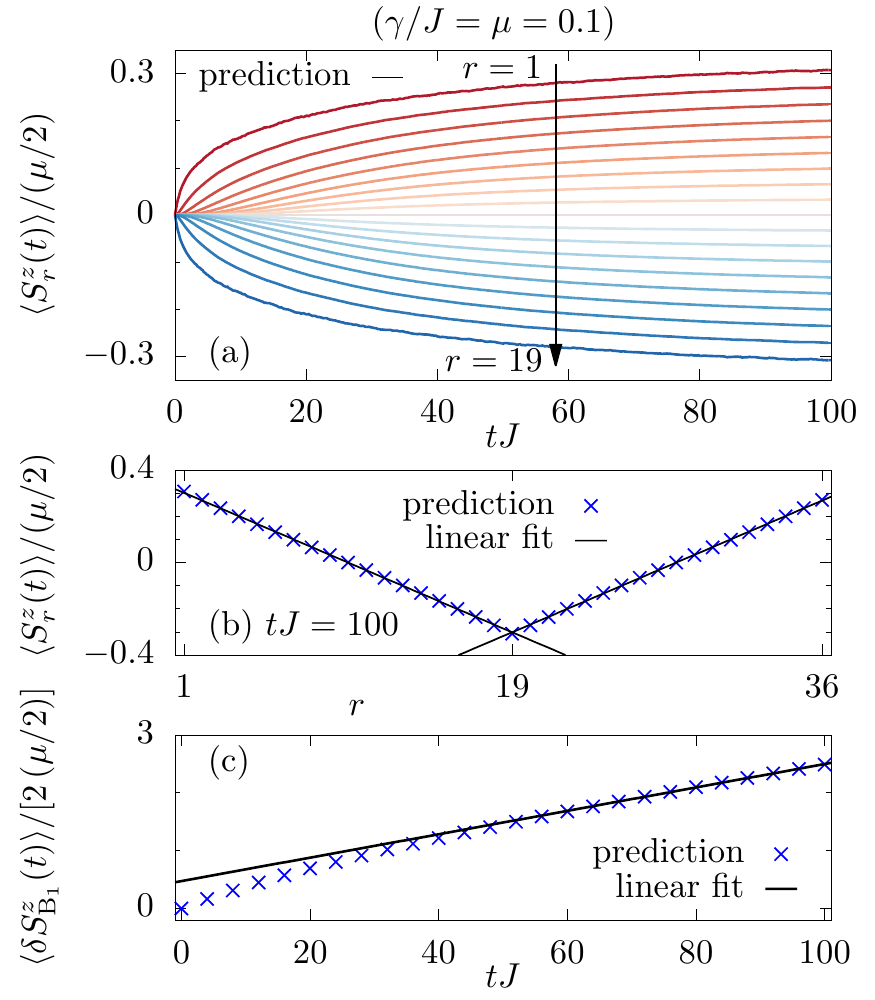}
  \caption{Prediction under the idealized assumption that the closed system is
  perfectly diffusive with a diffusion constant ${D/J = 0.6}$. 
  In the open system, 
  the corresponding diffusion constant obtained from the steady state is ${D/J 
\approx 0.60}$ \cite{D3}.
  }
  \label{fig:perfect_N36}
  \end{figure}

  \begin{figure}[tb]
    \centering
    \includegraphics[width=0.90\columnwidth]{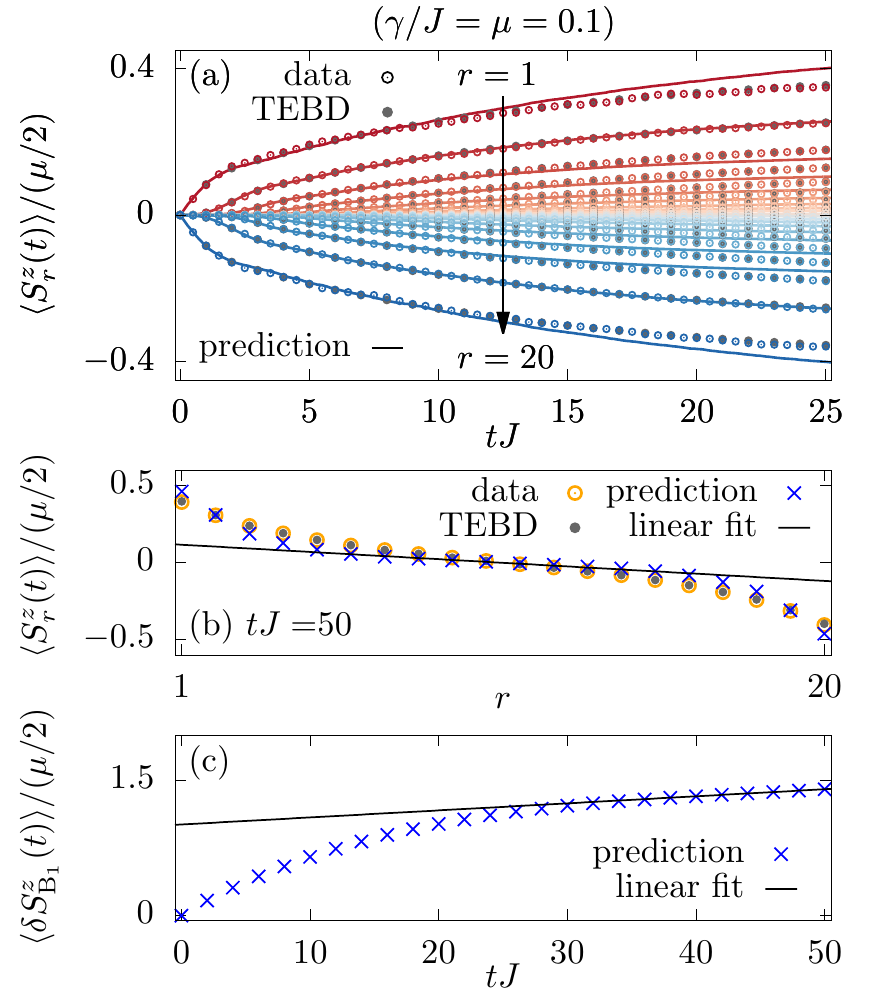}
    \caption{
    [(a) and (b)]  Analogous to Fig.~\ref{fig:comparison} of the main text, but 
  now with open boundary conditions and bath sites ${\BOne=1}$ and ${\BTwo=N}$, 
  where ${N=20}$. 
  Numerical results obtained by stochastic unraveling (data) are compared to 
our 
  prediction based on equilibrium correlation functions [cf. 
Eq.~\eqref{Eq::Average}]. 
  Furthermore, we show data obtained by 
  solving the Lindblad equation using a time-evolving block decimation (TEBD) 
  approach, which are in perfect agreement with the stochastic unraveling.   
  The diffusion constant extracted from the steady state is ${D/J \approx 
0.63}$~\cite{D15_OBC}.}
  \label{fig:D15_OBC}
    \end{figure}  
  
As discussed in Ref.\ \cite{Heitmann2022}, the central assumption within the 
above derivation is that the system has sufficient time to 
equilibrate between two jumps or, 
in other words, the magnetization injected at the contact 
sites has to spread over some region of the system. This requirement means 
that, in addition to a weak driving $\mu$, one has to choose a small 
coupling $\gamma$. Still, it might happen that even for a small coupling the 
equilibration process is hampered, as it is the case for open boundary 
conditions in certain models and parameter regimes, see the 
discussion below for more details.

\section*{Perfect Diffusion}

As mentioned in the main text, we attribute differences between diffusion
constants in open and closed systems to finite-size effects at long times,
where the steady state is established. To support this, it is instructive
to consider the idealized assumption of perfect diffusion in the closed system.
In this case, which has no finite-size effects at any time, the equilibrium
correlation functions take on the simple form \cite{Bertini2021}
\begin{equation}
\langle S_r^z(t) S_{r'}^z(0) \rangle_\text{eq}  = \frac{1}{4} \, e^{-2 
D_\text{closed}
t} \, {\cal I}_{r-r'}(2 D_\text{closed} t) \, ,
\end{equation}
where ${\cal I}_r(t)$ is the modified Bessel function of the first kind
and of the order $r$. Choosing ${D_\text{closed}/J = 0.6}$, we show our 
prediction for 
the open system in Fig.~\ref{fig:perfect_N36}. 
Indeed, using the relationship  ${D = -\langle j_r \rangle/\nabla \langle S_{r} 
\rangle}$, we find a
corresponding diffusion constant ${D/J \approx 0.60}$ in the steady state 
\cite{D3}. Thus, for perfectly diffusive behavior 
without finite-size effects, 
the steady state yields the same transport coefficient as the equilibrium 
correlation function (in 
contrast to the realistic case discussed in the context of Fig.\ 
\ref{fig:prediction_N36} in the main text).

Furthermore, as already discussed in the main text, 
it is possible to calculate $D(t)$ already at finite times before the 
steady state 
is established, via
\begin{equation}
D(t) = \frac{\partial_t \langle S_r^z(t) \rangle}{\nabla^2 \langle S_r^z(t) 
\rangle}
\end{equation}
with
\begin{equation}
\nabla^2 \langle S_r^z(t) \rangle = \langle S_{r-1}^z(t) \rangle - 2 \langle 
S_r^z(t) \rangle + \langle S_{r+1}^z(t) \rangle \, ,
\end{equation}
which is just the diffusion equation for a lattice in one spatial dimension.
Evaluating this expression for, e.g., site ${r = 9}$ and time ${t J = 25}$ 
for the perfectly diffusive data 
in Fig.~\ref{fig:perfect_N36}, we obtain the value
\begin{equation}
D/J \approx \frac{(0.012748 - 0.012062)/1}{0.025282 -2 \cdot 0.012062 + 0}
\approx 0.59 \, ,
\end{equation}
which is again in good agreement with $D_\text{closed}$.

\begin{figure}[tb]
  \centering
  \includegraphics[width=0.90\columnwidth]{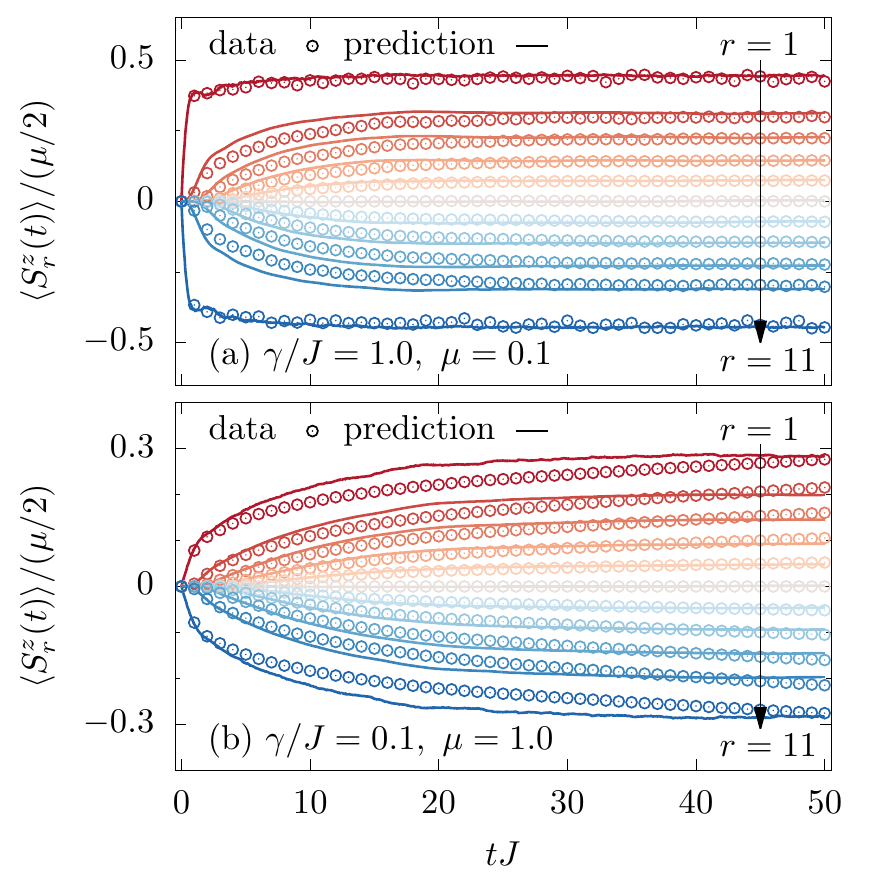}
  \caption{Time evolution of the local magnetizations 
  ${\langle S_r^z (t) \rangle}$ 
  for the model parameters in Fig.~\ref{fig:comparison}, but for the cases of 
  (a) 
  large coupling ${\gamma/J = 1}$ (${\mu = 0.1}$) and (b) strong driving ${\mu 
= 
1}$ 
  (${\gamma/J = 0.1}$). As before, the derived prediction 
  is compared to stochastic unraveling (data).}
  \label{fig:large_gamma}
  \end{figure}

\begin{figure}[b]
\centering
\includegraphics[width=0.90\columnwidth]{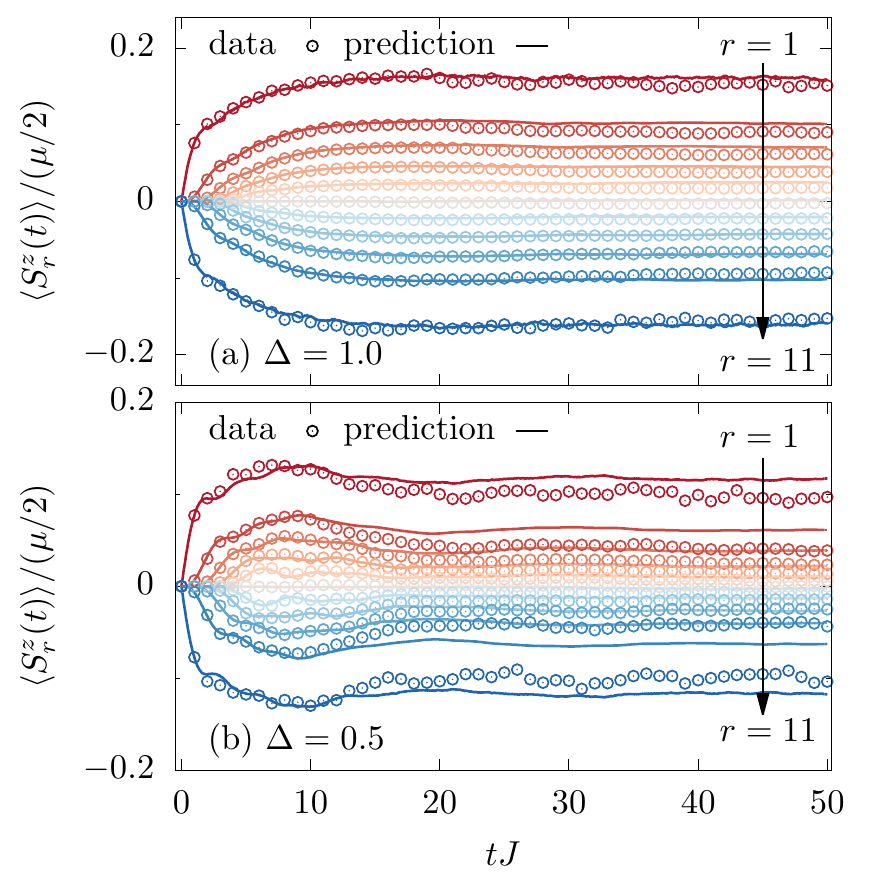}
\caption{Time evolution of the local magnetizations ${\langle S_r^z (t) 
\rangle}$ 
for the model parameters in Fig.~\ref{fig:comparison}, but for the 
anisotropies (a) ${\Delta = 1.0}$ and (b) ${\Delta = 0.5}$. As before, the 
derived 
prediction is compared to stochastic unraveling (data).}
\label{fig:Delta05}
\end{figure}  

\section*{Open Boundaries and comparison with TEBD simulations}

So far, we have focused on systems with 
periodic boundary conditions, which are the natural choice for 
closed systems, whereas 
state-of-the-art matrix product state approaches to open systems 
commonly rely on open boundary conditions with Lindblad driving at 
the systems' edges.

In Fig.~\ref{fig:D15_OBC}, we show a  
comparison between our theoretical prediction \eqref{Eq::Average} and full 
stochastic unraveling, similar to 
Fig.~\ref{fig:comparison} in the main text, but now for a XXZ chain 
with open 
boundaries, where 
the two baths are placed at the ends of the chain, ${\BOne=1}$ and ${\BTwo=N}$.
As shown in Fig.~\ref{fig:D15_OBC}(a), the time evolution of the local 
magnetization ${\langle S_r^z(t) \rangle}$
is well captured by the prediction 
\eqref{Eq::Average}, though deviations start to appear at times 
${tJ\gtrsim20}$.
These slight deviations might be 
caused by the fact that the equilibrium correlation functions can 
exhibit unusual behavior 
in the case of open boundary conditions.
In particular, 
$\langle S_{\BOne}^z\!(t) S_{\BOne}^z\!(0) \rangle$ and 
$\langle S_{\BTwo}^z\!(t) S_{\BTwo}^z\!(0) \rangle$ 
do not fully decay 
for any $\Delta>1$ in the case of open boundary conditions due to the 
presence of a strong zero mode, where the edge spins retain memory of their 
initial conditions for very long times \cite{Fendley2016,Kemp2017}. 

In addition to our prediction \eqref{Eq::Average} and the stochastic 
unraveling, we include in Fig.\ \ref{fig:D15_OBC} numerical data obtained 
by a state-of-the-art matrix product state implementation based on 
time-evolving block decimation (TEBD)
with time step $\mathrm{d}tJ=0.05$ and bond dimension $\chi=128$.
Importantly, we find that this TEBD data
is in perfect agreement with the results from stochastic unraveling. 

The convincing agreement between TEBD, stochastic unraveling, and our 
theoretical prediction is further highlighted in Fig.~\ref{fig:D15_OBC}(b),  
where the site dependence of the profile ${\langle S_r^z(t) \rangle}$ is 
shown for time ${tJ=50}$. Moreover, we find that this profile is well 
described by a linear function in the bulk, 
far away from the bath contacts.
The injected magnetization ${\langle \delta S^z_{\BOne}\!(t) \rangle}$  
is shown in Fig.~\ref{fig:D15_OBC}(c).
The diffusion constant in the open system is again given by  the ratio of
the slopes in Figs.~\ref{fig:D15_OBC}(b) and \ref{fig:D15_OBC}(c). 
This way, we find the value ${D/J \approx 0.63}$ 
in very good agreement with the value ${D/J \approx 0.6}$ in the closed system.
  
\section*{Injected magnetization}

The first Lindblad bath injects magnetization at the corresponding contact
site $r = \BOne$. This magnetization can also be predicted and written as
\begin{equation}
\langle \delta S^z_{\BOne}\!(t) \rangle \approx 
\frac{1}{\text{T}_\text{max}} \sum_{\text{T}=1}^{\text{T}_\text{max}} 
\delta 
d_{\BOne, \text{T}}(t)
\end{equation}
with 
\begin{equation}
\frac{\delta d_{\BOne,\text{T}}(t)}{2 \mu} = \sum_j A_j \, \Theta(t - 
\tau_j) \, \langle \,[S_{\BOne}^z\!(0)]^2 \rangle \, ,
\end{equation}
cf.\ Eq.~(\ref{eq:convolution}). This magnetization can then be related to
the local currents, which are all the same in the steady state, i.e.,
\begin{equation}
\langle j_r \rangle = \langle j_{r'} \rangle \, , \quad \BOne \leq r,r' \leq 
\BTwo \, .
\end{equation}
Thus, it is sufficient to know $\langle j_{\BOne} \rangle$, which follows from
the injected magnetization via
\begin{equation}
\langle j_{\BOne} \rangle = \frac{\text{d}}{\text{d} t}  \frac{\langle \delta
S^z_{\BOne}\!(t) \rangle}{2} \, ,
\end{equation}
where the factor $1/2$ takes into account that the injected magnetization can
flow to the left and to the right of this bath, due to periodic boundary
conditions. By the use of this 
expression, we find that, for the case discussed in Fig.~\ref{fig:comparison}, 
$\langle j_r\rangle/J\approx 0.0028$.

For comparison, we can calculate the local currents for the same model within 
the stochastic
unraveling as
\begin{equation}
  \langle j_r\rangle
  \approx 
  \frac{1}{\text{T}_\text{max}} \sum_{\text{T}=1}^{\text{T}_\text{max}}
  \frac{\bra{\psi_\text{T}(t)}j_r \ket{\psi_\text{T}(t)}}
  {\left\lVert |\psi_\text{T}(t) \rangle\right\rVert^2}
\end{equation}
for $\mathcal{O}(10^4)$ trajectories. For long times $t J \gtrsim 30$,  where
the steady state is established,
we find a corresponding value of $\langle j_r\rangle/ J \approx 0.003$,
which is close to the predicted one above.

\section*{Large Coupling / Strong Driving}

In the main text, we have focused on the case of small coupling 
${\gamma/J = 0.1}$ 
and weak driving ${\mu = 0.1}$, where we have found a convincing 
agreement between the derived prediction and exact numerics in 
Fig.~\ref{fig:comparison}. 
To illustrate that deviations occur for larger coupling 
or stronger driving, we depict a corresponding 
comparison for (a) ${\gamma/J = 1}$ (${\mu = 0.1}$) and 
(b) ${\mu = 1}$ (${\gamma/J = 0.1}$) in Fig.~\ref{fig:large_gamma}. 
In both cases (a) and (b), deviations are visible already at finite 
times before the steady state is reached. Interestingly, in the case (a) of 
strong coupling, the overall agreement is still satisfactory and the profile in 
the steady state is predicted accurately.

\section*{Other Anisotropies}

In Fig.~\ref{fig:comparison} of the main text, 
we have provided a comparison of the magnetization dynamics for
anisotropy ${\Delta = 1.5}$. 
Complementarily, we show a comparison for anisotropies 
(a) ${\Delta = 1.0}$ and 
(b) ${\Delta = 0.5}$ in Fig.~\ref{fig:Delta05}. 
For the case of ${\Delta = 1.0}$, 
we again find a convincing 
agreement. While for the case of ${\Delta = 0.5}$ the overall behavior of
prediction and numerics is still similar, deviations are visible at
long time scales. These deviations reflect that the prediction is not
exact in a mathematical sense but involves physical assumptions (such as,
e.g., on the equilibration properties of the involved 
equilibrium correlation functions)
which may not hold perfectly in any given situation.

\balancecolsandclearpage

\end{document}